\documentclass[aps,prd,amssymb,groupedaddress]{revtex4}
\usepackage{epsfig}
\usepackage[english]{babel}

\usepackage{amsfonts}
\usepackage{amssymb}
\usepackage{amsmath}
\usepackage{multirow}

\def\a{\alpha}
\def\b{\beta}

\def\d{\delta}

\def\l{\lambda}

\def\W{\Omega}

\def\z{\zeta}

\def\pl{\parallel}
\def\pp{\bot}
\def\px{\approx}

\def\={\nonumber &=}
\def\nn{\nonumber}
\def\&{{}&}

\def\({\left(}
\def\){\right)}
\def\[{\left[}
\def\]{\right]}
\def\<{\left\langle}
\def\>{\right\rangle}

\def\ux{{\bf \hat{x}}}
\def\bk{{\bf k}}

\def\curl{\mathcal}

\def\eq{\begin{eqnarray}}
\def\qe{\end{eqnarray}}

\def\and{\quad \mbox{and} \quad}

\def\fnl{f_\textrm{NL}}

\def\Fnl{ F_\textrm {NL}}

\def\bfnl{\kern2pt\overline{\kern-2ptf}_\textrm{NL}}

\def\lmax{l_\textrm{max}}

\def\blll{b_{l_1l_2l_3}}
\def\hlll{h_{l_1l_2l_3}}

\def\barQ{\kern2pt\overline{\kern-2pt\curl{Q}}}

\def\barR{\kern2pt\overline{\kern-2pt\curl{R}}}

\def\nmax{n_\textrm{max}}

\def\setsize{\csname @setfontsize\endcsname \setsize}

\begin{document}


\title{Optimal Polyspectra Estimation}

\author{J.R.~Fergusson}

\author{E.P.S.~Shellard}

\affiliation
{Centre for Theoretical Cosmology,\\
Department of Applied Mathematics and Theoretical Physics,\\
University of Cambridge,
Wilberforce Road, Cambridge CB3 0WA, United Kingdom}

\date{\today}

\begin{abstract}

{\setsize{9}{12}
\noindent We show that full inverse covariance weighting can be naturally incorporated 
into modal estimation methods making them optimal for  the CMB power spectrum, bispectrum and trispectrum, as well as in other 
3D applications.   Modal estimation methods are highly efficient requiring inversion of only an $\nmax \times \nmax$ covariance matrix,
where $\nmax$ is the limited number of modes needed to describe both the theoretical model and the noise which 
affects it (in contrast to the  full $\lmax^2 \times \lmax^2$ inverse covariance weighting).     
For an isotropic modal bispectrum estimator implemented at a WMAP resolution, we demonstrate how the modal inverse covariance weighting improves optimality significantly. We describe optimal modal estimation in general terms applicable to polyspectra of any order in a  formalism that can 
be efficiently extended beyond the isotropic case.   We also discuss further improvements 
 by characterising and preconditioning  the covariance matrix using a single bispectrum variance shape induced by the noise and mask. 
  We use the example of anisotropic noise in bispectrum estimation to illustrate these methods and demonstrate the necessity of subtracting a linear estimator term for optimality, removing cross-terms in
the variance.   We briefly discuss applications to the CMB power spectrum and trispectrum,  as well as  large-scale structure.  
}
\end{abstract}


\maketitle


\setsize{11}{13}

\section{Introduction}

A separable eigenmode approach to CMB bispectrum and trispectrum estimation has been developed
and implemented \cite{Fergusson:2006pr,FergussonLiguoriShellard2009}, providing a reconstruction of the WMAP CMB bispectrum and a wide variety of constraints on 
theoretical bispectra \cite{Fergusson:2010dm} and trispectra \cite{Fergusson:2010gn}.  
In this paper, we demonstrate how this modal methodology can be made fully optimal by incorporating inverse covariance weighting
in the presence of anisotropic noise and other systematic effects. It   can be efficiently implemented 
provided the eigenmode expansion utilised has sufficient resolution to describe, first,  the theoretical 
polyspectrum shape being investigated and, secondly, the noise, mask and other systematic contributions that are correlated with it.   
The generality and simplicity of this modal covariance inversion, combined with the 
efficiency of the separable modal estimator, opens up the possibility of fully optimal analysis of high resolution CMB experiments 
such as the Planck satellite, as well as higher order correlators in  the  three-dimensional distribution of  large-scale structure. 

Full inverse covariance weighting for CMB power spectrum analysis has a long history going back to the study of the 
COBE data using direct Cholesky decomposition with  ${\cal O}(\lmax^6)$ operations (see, for example, \cite{Gorski:1994ye, Bond:1994aa}).
This was not feasible for higher resolution experiments such as WMAP, but could be improved to ${\cal O}(\lmax^4)$ with conjugate gradient 
methods and iterative approximations  \cite{Oh:1998sr}.   The importance of inverse covariance weighting for the CMB bispectrum 
has also been discussed for some time (see, for example, \cite{9804222}) but was only tackled at WMAP resolution more recently using multigrid preconditioning and conjugate gradient methods \cite{Smith:2007rg,SmithetAl2009}.  Achieving full inversion at Planck resolution remains formidable goal for a realistic noise model with non-trivial correlations (see, for example, ref.~\cite{Wandelt2012}).   The significant advance of  the present work 
is that we show that it is not necessary to invert the full covariance matrix to obtain an optimal variance.  Instead, we invert a  much smaller 
modal covariance matrix from a subspace with sufficient degrees of freedom to describe the theoretical model, as well as the masking and noise modes
that actually contaminate it.  For example, we will investigate an isotropic bispectrum $\blll$ affected by anisotropic 
instrument noise or a cut-sky mask, discussing the projection of the 
pixel-variance noise map, variance and bispectrum which can affect the estimation.   We have previously demonstrated, in the isotropic case, that a highly efficient and general modal bispectrum estimator can be implemented
projecting the CMB data into a much smaller  subspace spanned by the relevant bispectrum modes \cite{Fergusson:2006pr,FergussonLiguoriShellard2009}.      
What is remarkable is the small number of modes required to characterise almost all popular theoretical bispectrum models
 $\nmax = {\cal O}(10-100)$  \cite{Fergusson:2008ra}.  For this isotropic subspace, we then noted that the required mode number is not noticeably increased when characterising the relevant noise and mask effects \cite{FergussonLiguoriShellard2009}.   Our expectation, therefore, when seeking an isotropic theoretical polyspectra estimator on the full anisotropic space that is that we will require  few additional modes to incorporate relevant noise and masking effects (this will be the subject of a future publication).    
Returning to the full isotropic implementation in a WMAP-realistic context, we are able to significantly improve the optimality of the variance using a much simpler inverse modal covariance weighting, getting about 30\% closer to that predicted by a Fisher matrix analysis than previous isotropic bispectrum estimators.

Finally we comment on other modal estimators such as wavelets \cite{Curto:2010si}, bins \cite{Bucher:2009nm} and needlets \cite{Rudjord:2009mh} which differ in there choice of initial basis but are otherwise identical.
Important distinctions remain, including the ability to analytically calculate modes, the number of modes required for convergence and the ability to project between primordial and CMB(or LSS)-spectra. 
Early versions of the wavelet method\cite{Curto:2010si}, claimed optimality while neglecting to incorporate the linear term in the estimator which is required to minimise the variance \cite{CreminellietAl2006}, though this has now been incorporated \cite{Curto:2012}.
We demonstrate the crucial role of this linear term for optimal estimation in the realistic situation where rotational invariance is broken.   

The plan of the paper is as follows.   We will first outline a general polyspectrum estimation methodology describing projections from 
the full polyspectrum space into the modal subspace of interest. In this context we show how the full covariance matrix is projected 
down into a much smaller modal covariance matrix which can be easily inverted.  Next we discuss the modal bispectrum 
estimator in detail, deriving the relationship between the full and modal covariance matrices.  WMAP results are presented 
comparing analytic optimal Fisher matrix forecasts with the variance obtained from Gaussian 
simulations for the general isotropic modal estimator.  We calculate the bispectrum variance from anisotropic noise and we note that 
identifying the primary bispectrum 
noise shape contributing to this variance can be used to efficiently precondition the covariance matrix for inversion. 
Finally, we give the modal estimation components required for optimal CMB power spectrum  and trispectrum analysis and briefly discuss 
large-scale structure correlators.

\section{Modal CMB polyspectra estimation}

The CMB polyspectrum $\langle\mathfrak{a}_{\wp}\rangle$ of degree $p$ for a given non-Gaussian model 
is the expectation value of the product of $p$ temperature multipole $a_{lm}$'s obtained
from an ensemble of universes, that is, 
\begin{align}
\langle \mathfrak{a}_{\wp}\rangle ~\equiv~ \langle a_{l_1m_1} a_{l_2m_2} ...a_{l_pm_p}\rangle 
\end{align}
where the index $\wp$ represents  the 2$p$ degrees of freedom $\wp =\{l_1,m_1, l_2,m_2, ..., l_p,m_p\}$ which 
cover the full domain ${\cal V}$ of possible polyspectra (up to a given resolution $l\le \lmax$). Here we assume $\langle \mathfrak{a}_{\wp}\rangle$ is the connected part of the correlator we seek. Most primordial theories predict an isotropic polyspectrum ${a}_{\ell}$ which resides on a smaller
$d$-dimensional subspace ${\cal V}_{\rm I}$, after integrating out (or summing over) the irrelevant anisotropic modes.    The 
isotropic ${a}_{\ell}$ has an equivalent counterpart in the full space $\cal V$ found 
by multiplying by  ${\cal G}_ {\ell\,\ell^\perp}$
\begin{align}
\<\mathfrak{a}_{\wp}\> = {\cal G}_ {\ell\,\ell^\perp}\,\<a_{ \ell}\>\,,
\end{align}
where $\wp=\{\ell,\ell^\perp\}$ with the $\ell$ index covering the $d$ isotropic degrees of freedom and $\ell^\perp$ 
the remaining $2p$$-$$d$ anisotropic degrees.   For concreteness, we note that the power spectrum $C_l$ has $d=1$ with   ${\cal G}_{\ell\,\ell^\perp} = \int d\W_x Y_{l_1 m_1}(\ux) Y^*_{l_2 m_2}(\ux) = \delta _{l_1l_2}\delta_{m_1 m_2}$, as we work with $a_{lm}a^*_{lm}$, while for the bispectrum we have $d=3$ with the Gaunt integral   
${\cal G}_{\ell\ell^\perp}= \curl{G}^{l_1 l_2 l_3}_{m_1 m_2 m_3} 
\equiv \int d\W_x Y_{l_1 m_1}(\ux) Y_{l_2 m_2}(\ux) Y_{l_3 m_3}(\ux),$  $\;\ell \rightarrow \{l_1,l_2,l_3\},\; \ell^\perp\rightarrow \{m_1,m_2,m_3\}$
(and similar for higher `nondiagonal' polyspectra).  

When investigating CMB polyspectra we are typically 
seeking to match a statistically isotropic prediction $\mathfrak{ a}_{\wp_i}$ to the CMB data, 
with an appropriate weighting to reflect our knowledge of the signal-to-noise
for the specific experiment.   The estimator $\bar{\cal E}$ determines the goodness of fit between 
these through the inner product  
\begin{align}
\bar{\cal E}\; \equiv\; \frac{\sum _{\wp\,\wp'} \langle \mathfrak{a}_{\wp} \rangle \mathfrak{C}^{-1}_{\wp \wp'} \(\mathfrak{a}_{\wp} - \mathfrak{a}^{lin}_{\wp}\)}{\sum _{\wp\,\wp'} \langle \mathfrak{a}_{\wp}\rangle \mathfrak{C}^{-1}_{\wp \wp'} \langle \mathfrak{a}_{\wp}\rangle}\,,
\end{align}
where the $\mathfrak{C}^{-1}_{\wp \wp'} \equiv C^{-1}_{l_1m_1,l'_1 m'_1}... C^{-1}_{l_p m_p,l'_p m'_2p}$ represents the inverse covariance across all $p$ multipoles.  Here, $\mathfrak{a}^{lin}_{\wp}$ represents the `linear term' used to subtract out known systematic effects; for the power spectrum, $\mathfrak{a}^{lin}_{\wp} = C_{l_1m_1,l_2,m_2}$ and for the bispectrum 
$\mathfrak{a}^{lin}_{\wp} = 3C_{l_1m_1,l_2,m_2} a_{l_3 m_3}$.  In practice, the linear term is calculated from Monte Carlo simulations of Gaussian maps with the correct experimental effects (mask, beam, noise) included, see \cite{FergussonLiguoriShellard2009}.  We also note that $\langle \mathfrak{a}_{\wp}\rangle$ is understood to be the theoretical prediction modified to include the relevant experimental effects.

If we define the weighted vectors $\cal A ,\,B$ and the matrix $\cal C$
\begin{align} \label{eq:legoestimator}
{\cal A}_{\wp} = \frac{\langle \mathfrak{a}_{\wp} \rangle}{\sqrt{C_{l_1}C_{l_2} ... C_{l_p}}} \,,\qquad
{\cal B}_{\wp} = \frac{\mathfrak{a}_{\wp} - \mathfrak{a}^{lin}_{\wp}}{\sqrt{C_{l_1}C_{l_2} ... C_{l_p}}} \,,\qquad
{\cal C}_{\wp \wp'} = \frac{\mathfrak{C}_{\wp \wp'}}{\sqrt{C_{l_1}C_{l'_1} ... C_{l_p}C_{l'_p}}}\,,
\end{align}
then the estimator can be written in matrix notation as
\begin{align}\label{eq:stdestimator}
\bar{\cal E} = \frac{{\cal A}^T {\cal C}^{-1} {\cal B}}{{\cal A}^T {\cal C}^{-1} {\cal A}}\,.
\end{align}
Modal decompositions have proven to be very useful for estimating general CMB bispectra and trispectra. Suppose we have a basis ${\cal R}_{n \wp}$ where n is the mode number. We will take it to be orthonormal so
\begin{align}
\sum_{\mathfrak p} {\cal R}_{n {\mathfrak p}}  {\cal R}_{n' {\mathfrak p}}  = \d_{n n'}\,,
\end{align}
or in matrix notation ${\cal R} {\cal R}^T = I$. We can then represent our theoretical polyspectra in this basis using mode coefficients $\a$ defined by
\begin{align}
{\cal A}_{\mathfrak p} = \sum_n \a_n R_{n {\mathfrak p}} \quad \({\cal A} =  {\cal R}^T \a\)\,.
\end{align}
The $\a$ can then be calculated easily via $\a = {\cal R}{\cal A}$.   

Most theoretical cosmological models have rotational invariance at their foundation, so the polyspectra ${\cal A}^{\rm th}$ they predict are generically isotropic.   For this reason, our starting point must be to include sufficient isotropic mode functions in our basis to accurately characterise them. Most theoretical bispectra  ${\cal A}^{\rm th}$ have smooth or simple forms and so for a suitable choice of ${\cal R}$ they can in fact be represented by surprisingly few of these isotropic modes (for example, most scale-invariant bispectrum models require only 15 modes at WMAP resolution \cite{Fergusson:2008ra}). This motivates us to work with a truncated basis containing $n_{max}$ modes which span the space of interest which is determined by the class of theoretical models under consideration modified by experimental effects, i.e.\ masking and noise.  For an isotropic bispectrum estimator, we have shown that the experimental effects of the mask and noise can be incorporated without increasing $n_{max}$ substantially \cite{FergussonLiguoriShellard2009}.  For anisotropic contributions from experimental effects which yield a partially off-diagonal  ${\cal A}$, we need to supplement our basis functions with additional anisotropic modes.  In line with the remarkable data compression possible in the isotropic case, we anticipate only increasing $n_{max}$ by multiplying by a factor of a few provided the extra anisotropic ${\cal R}$ are carefully chosen.    

Given that $n_{max}$ is so small relative to the vast number $2p$ of multipole modes, the ${\cal R}$ become highly rectangular $n_{max}$$\times 2p$ matrices.  We can now define a projection operator from the  full space $\cal V$ to the subspace ${\cal V}_{\cal P}$ 
\begin{align}
{\cal P} \equiv {\cal R}^T {\cal R}
\end{align}
Here, we assume we have truncated our basis such that our theory and distorting experimental effects remain accurately described, ie. $\cal PA = A$. Using this projection we define the modal counterparts of the data on the same subspace ${\cal V}_{\cal P}$,
\begin{align}
 \b = {\cal R}{\cal B} &\longrightarrow {\cal PB} =  {\cal R}^T \b\,,
 \end{align}
 as well as the covariance
 \begin{align}
\zeta \;= \;{\cal R}{\cal C}{\cal R}^T\,.
\end{align}
The projected covariance is related to the covariance of $\b$ by $p!$ as we now show.
\begin{align}
\<\b \b^T\> &= \mathcal{R}\<\mathcal{B}\mathcal{B}^T\>\mathcal{R}^T \\
&= \mathcal{R} \frac{ \< \mathfrak{a}_{\wp} \mathfrak{a}^{\vphantom{l}}_{\wp'} \> - \< \mathfrak{a}_{\wp} \mathfrak{a}^{lin}_{\wp'} \> - \< \mathfrak{a}^{lin}_{\wp} \mathfrak{a}_{\wp'} \> + \< \mathfrak{a}^{lin}_{\wp} \mathfrak{a}^{lin}_{\wp'} \> }{\sqrt{C_{l_1} ... C_{l_p}C_{l'_1} ... C_{l'_p}}} \mathcal{R}^T\,.
\end{align}
As we know the $a_{lm}$ are very close to Gaussian we can expand $\< \mathfrak{a}_{\wp} \mathfrak{a}_{\wp'} \>$ into the product $\<a_{lm} a_{l'm'}\>$,$\<a_{lm} a_{lm}\>$ and $\<a_{l'm'} a_{l'm'}\>$. In general there will be $p!$ ways to have all pairs consist of one primed and one unprimed $a_{lm}$ plus $(2p)!/2p! - p!$ combinations with at least one pairing between two unprimed and two primed $a_{lm}$. The linear term is always chosen such that it removes all contributions of the second type. As a result we have $\<\mathcal{B}\mathcal{B}^T\> = p! \mathcal{C}$ and so
\begin{align}
\zeta \;= \;{\cal R}{\cal C}{\cal R}^T = \frac{1}{p!}\<\b \b^T\>
\end{align}
Using these quantities we can define a new estimator 
\begin{align}\label{eq:modalestimator}
{\cal E} ~ &\equiv ~\frac{\a^T \zeta^{-1} \b}{\a^T \zeta^{-1} \a} \\
&= ~\frac {({\cal RA})^T{\cal R}{\cal C}^{-1} {\cal R}^T {\cal RB}} {{\cal RA}^T {\cal R}{\cal C}^{-1} {\cal R}^T  {\cal RA} } \nn\\ &= \frac {{\cal A}{\cal P} {\cal C}^{-1} {\cal P} {\cal B}} {{\cal A}^T{\cal P} {\cal C}^{-1} {\cal P}{\cal A} }
\end{align}
where we have used $({\cal R}{\cal C}{\cal R}^T)^{-1} = {\cal R}{\cal C}^{-1}{\cal R}^T + Z_\pp$ where $Z_\pp$ is an arbitrary term orthogonal to the projection which we can ignore. Strictly $\cal C$ is non-invertible, formally because it contains duplicate data due to permutations, but also because of loss of information from a cut-sky mask.   However, this problem can be circumvented.   For example, we can restrict $\wp$ to count only combinations where $l_1 \le l_2 \le l_3$, with a suitable restriction on $m_i$ when the $l_i$ are equal, and weight them appropriately to remove this complication. Once it is inverted we can then reinsert duplicates to return to the previous ordering for $\wp$.  Given the estimator expression \eqref{eq:modalestimator}, it can be easily seen that its variance is given by  $\sigma^2 \propto (\a^T \zeta^{-1} \a)^{-1}$ since $\zeta \propto \langle \b\b^T\rangle$.

We now wish to determine how closely the standard estimator \eqref{eq:stdestimator} and the modal estimator \eqref{eq:modalestimator} coincide. In the idealised case of homogeneous noise and no mask the covariance matrix is diagonal and so ${\cal C} = I$ and they are clearly identical. Also in the limit of completeness $n_{max}$$\rightarrow {2 p}$, we have ${\cal P} \rightarrow I$ and again the estimators are equivalent. In a realistic context with inhomogeneous noise, a mask and a truncated set of modes the difference can be quantified by considering
\begin{align}
{\cal A} = \[\begin{array}{c} {\cal A}_\pl \\ 0 \end{array}\] \qquad {\cal B} = \[\begin{array}{c} {\cal B}_\pl \\ {\cal B}_\pp \end{array}\] \qquad {\cal C}^{-1} = \[\begin{array}{cc} {\cal C}^{-1}_\pl & {\cal C}^{-1}_\times \\{{\cal C}^{-1}}^T_\times &  {\cal C}^{-1}_\pp \end{array}\]
\end{align}
where we define the vector ${\cal X}_\pl \equiv {\cal PX}$ to be the component of ${\cal X}$ in the subspace ${\cal V}_{\cal P}$ and the vector ${\cal X}_\pp \equiv (I-{\cal P}){\cal X}$ to be orthogonal to it and, similarly, for the matrix we can decompose it into ${\cal M}_\pl \equiv {\cal PMP}$, ${\cal M}_\pp \equiv (I-{\cal P}){\cal M}(I-{\cal P})$ and ${\cal M}_\times \equiv {\cal P}{\cal M}(I-{\cal P})$. In this case, we can represent the standard estimator (\ref{eq:stdestimator}) in the form,
\begin{align}
\bar{{\cal E}} = \frac {{\cal A}_\pl \( {\cal C}_\pl^{-1}{\cal B}_\pl + {\cal C}_\times^{-1}{\cal B}_\pp\)} 
{{\cal A}_\pl^T {\cal C}_\pl^{-1} {\cal A}_\pl }\,,
\end{align}
whereas our new modal estimator (\ref{eq:modalestimator}) is
\begin{align}\label{eq:newmodalestimator}
{\cal E} = \frac {{\cal A}_\pl{\cal C}_\pl^{-1}{\cal B}_\pl} 
{{\cal A}_\pl^T {\cal C}_\pl^{-1} {\cal A}_\pl }\,.
\end{align}
The difference between the estimators is the inverse covariance projection of the perpendicular part of the data onto the parallel space. We should however pause to note that the normalisation in both cases is identical. As the variance we expect for the estimator is the reciprocal of the normalisation (with a factor for permutations), we can see the effect of the cross-projection by comparing the variance we obtain from simulations with this prediction. If they differ we must deduce that our covariance weighting is sub-optimal and the cross-term is important in which case we must increase $n_{max}$ and expand our basis to include the relevant parts.  

It should be noted that as the covariance matrix $\z$ is symmetric positive definite it is possible to find its Cholesky decomposition,
\begin{align}
\zeta = \tilde\lambda\,\tilde \lambda^T\,.
\end{align}
where $\tilde\lambda$ is lower triangular. We can then absorb these factors into our $\a$ and $\b$,
\begin{align}\label{eq:Cholrot}
\a' =\tilde \l ^{-1} \a \quad \b' =\tilde \l ^{-1} \b\,.
\end{align}
This is equivalent to re-orthonormalising our basis with respect to the full covariance matrix, and hence the $\b$ are now uncorrelated. This is obviously a desirable property and has several applications which we explore elsewhere in future publications. The most obvious is that the estimator takes a very simple form
\begin{align}
{\cal E} = \frac{\a'^T \b'}{\a'^T \a'}
\end{align}
as by construction $\z' = I$.   

The computational advantages of the modal estimator (\ref{eq:modalestimator}) are considerable, especially when we note that 
direct evaluation of the full optimal estimator (\ref{eq:stdestimator}) is extremely challenging at high resolution. 
This is because, first, it represents a huge sum over $\lmax^{\,2p}$ quantities and, secondly, 
the inverse covariance weighting requires a huge $\lmax^2 \times \lmax^2$ matrix to be calculated and then inverted. As we noted in the introduction, there 
do exist methods which make this tractable in specific cases, however for large $\lmax$ and general noise models at Planck resolution this has not yet 
been achieved.
Calculation of the mode coefficients $\a,\b$ may appear to be troublesome, but it is made highly efficient using a separable form which has been discussed
at length previously \cite{FergussonLiguoriShellard2009}. 
Instead, here, we focus on the improvements offered by the modal approach for inverse covariance weighting, reducing this 
to the inversion of a small $\nmax \times\nmax$ matrix. 

\section{CMB bispectrum estimation}

\subsection{Inverse covariance weighting with the modal bispectrum estimator}

Using the CMB bispectrum, we will now demonstrate a concrete example and implementation of modal polyspectra estimation with inverse covariance weighting.  The estimator for the bispectrum takes the general form
\begin{align}\label{eq:bispectrumest}
{\cal E} = \frac{1}{N} \sum_{l_i m_i} B^{\,l_1 \,l_2 \,l_3}_{m_1 m_2 m_3}  C^{-1}_{l_1 l'_1 m_1 m'_1}  C^{-1}_{l_2 l'_2 m_2 m'_2}  C^{-1}_{l_3 l'_3 m_3 m'_3} \(a_{l'_1 m'_1} a_{l'_2 m'_2} a_{l'_3 m'_3} - 3 C_{l'_1 l'_2 m'_1 m'_2} a_{l'_3 m'_3} \)\,,
\end{align}
where  $B^{l_1 l_2 l_3}_{m_1 m_2 m_3} = \langle a_{l_1 m_1} a_{l_2 m_2} a_{l_3 m_3}\rangle$ is the full bispectrum, modified from being purely isotropic by experimental effects, and $N$ is the appropriate normalisation. This estimator has been shown to be optimal in the ideal case without mask or noise in the absence of a linear term \cite{0503375}.   The linear term was proposed
in \cite{CreminellietAl2006} to minimise the variance in a realistic setting when rotational invariance is broken. To construct the modal bispectrum estimator, the quantities  $\cal A,B,C$ in (\ref{eq:legoestimator}) take  the form
\begin{align}
{\cal A} &=  \frac{B^{\,l_1 \,l_2 \,l_3}_{m_1 m_2 m_3}}{\sqrt{C_{l_1}C_{l_2}C_{l_3}}}\,,\\
{\cal B} &= \frac{a_{l_1 m_1}a_{l_2 m_2}a_{l_3 m_3} - 3\,C_{l_1 m_1 ,l_2 m_2}a_{l_3 m_3}}{\sqrt{C_{l_1}C_{l_2}C_{l_3}}}\,,\\
{\cal C} &= \frac{C_{l_1 m_1 ,l'_1 m'_1}C_{l_2 m_2 ,l'_2 m'_2}C_{l_3 m_3 ,l'_3 m'_3}}{\sqrt{C_{l_1}C_{l_2}C_{l_3}C_{l'_1}C_{l'_2}C_{l'_3}}}\,.
\end{align}
For simplicity and to relate the discussion more easily to previous work, let us restrict attention to an isotropic subspace, though the expressions derived below can be converted for the general anisotropic case.   We shall take the usual approximation that the primary contribution to the isotropic bispectrum is the predicted theoretical model itself, that is,  
\begin{align}
{\cal A} &= \curl{G}^{\, l_1\, l_2\, l_3}_{m_1 m_2 m_3} \frac{b_{l_1 l_2 l_3}}{\sqrt{C_{l_1}C_{l_2}C_{l_3}}}\,
\end{align}
where $b_{l_1 l_2 l_3}$ is the predicted reduced bispectrum. Here we note that we are always working with quantities that include the relevant experimental effects. For example, as we are only considering the isotropic case we can approximate $b_{l_1 l_2 l_3} \px f_{sky} b_{l_1} b_{l_2} b_{l_3} b^{clean}_{l_1 l_2 l_3}$ and $C_l \px f_{sky}\( b_l^2 C^{clean}_l + N_l\)$ where $b_l$ is the beam window function, $N_l$ is the noise power spectrum and $f_{sky}$ is the fraction of the sky remaining after masking.

An isotropic set of basis functions can be represented in the full anisotropic bispectrum space by
\begin{align}\label{eq:isobasisbi}
{\cal R}^{\,l_1 \,l_2 \,l_3}_{m_1 m_2 m_3} = \frac{\curl{G}^{l_1 l_2 l_3}_{m_1 m_2 m_3}}{v_{l_1}v_{l_2}v_{l_3}} R_{n l_1 l_2 l_3}\,,
\end{align}
where $v_l$ is a weight function chosen to mimic the scaling of $\curl{G}$, the Gaunt integral.  The simple orthogonality condition ${\cal R} {\cal R}^T = I$ in the full space ${\cal V}$, yields  a weighted orthogonality condition on $R$ in the projected isotropic subspace ${\cal V}_{\cal P}$,
\begin{align}
\d_{nn'} = \sum_{l_i m_i}\frac{\(\curl{G}^{l_1 l_2 l_3}_{m_1 m_2 m_3}\)^2}{v^2_{l_1}v^2_{l_2}v^2_{l_3}} R_{n l_1 l_2 l_3} R_{n' l_1 l_2 l_3} = \sum_{l_i}\frac{h_{l_1l_2l_3}^2}{v^2_{l_1}v^2_{l_2}v^2_{l_3}} R_{n l_1 l_2 l_3} R_{n' l_1 l_2 l_3}
\end{align}
where $h_{l_1l_2l_3}$ is a geometric factor defined by
\begin{align}
h_{l_1l_2l_3} = \sqrt{\frac{(2l_1+1)(2l_2+1)(2l_3+1)}{4 \pi }} \(\begin{array}{ccc} l_1 & l_2 & l_3 \\ 0 & 0 & 0 \end{array}\)
\end{align}
and decomposing the  $\cal A,B,C$ into mode space yields the $\alpha_n,\,\beta_n$ coefficients and the covariance matrix $\zeta_{nn'}$ respectively, that is, 
\begin{align}\label{eq:abispect}
\a_n &= \sum_{l_i}\frac{(2l_1+1)(2l_2+1)(2l_3+1)}{4 \pi v^2_{l_1}v^2_{l_2}v^2_{l_3}} \(\begin{array}{ccc} l_1 & l_2 & l_3 \\ 0 & 0 & 0 \end{array}\)^2 \frac{v_{l_1}v_{l_2}v_{l_3} b_{l_1 l_2 l_3}}{\sqrt{C_{l_1}C_{l_2}C_{l_3}}} R_{n l_1 l_2 l_3} \\ \label{eq:bbispect}
\b_n &= \sum_{l_i m_i}\frac{\curl{G}^{l_1 l_2 l_3}_{m_1 m_2 m_3}}{v_{l_1}v_{l_2}v_{l_3}} \frac{a_{l_1 m_1}a_{l_2 m_2}a_{l_3 m_3} - 3\,C_{l_1 m_1 ,l_2 m_2}a_{l_3 m_3}}{\sqrt{C_{l_1}C_{l_2}C_{l_3}}}  R_{n l_1 l_2 l_3} \\ \label{eq:cbispect}
\z_{nn'} &= \sum_{l_i m_i l'_i m'_i}\frac{\curl{G}^{l_1 l_2 l_3}_{m_1 m_2 m_3}\curl{G}^{l'_1 l'_2 l'_3}_{m'_1 m'_2 m'_3}}{v_{l_1}v_{l_2}v_{l_3}v_{l'_1}v_{l'_2}v_{l'_3}} R_{n l_1 l_2 l_3} \frac{C_{l_1 m_1 ,l'_1 m'_1}C_{l_2 m_2 ,l'_2 m'_2}C_{l_3 m_3 ,l'_3 m'_3}}{\sqrt{C_{l_1}C_{l_2}C_{l_3}C_{l'_1}C_{l'_2}C_{l'_3}}} R_{n' l'_1 l'_2 l'_3}
\,.\end{align}
Calculation of $\a_n,\, \b_n$ is straightforward using methods described in detail in ref.~\cite{FergussonLiguoriShellard2009} and, as we have chosen the linear term correctly to remove cross terms in the variance, $\zeta = 1/6 \<\b \b^R\>$ as we demonstrate below
\begin{align}\label{eq:linearterm}
\nn\langle \beta_n\,\beta_{n'}\rangle &= \sum_{l_i m_i l'_i m'_i}\<\(\frac{\curl{G}^{l_1 l_2 l_3}_{m_1 m_2 m_3}}{v_{l_1}v_{l_2}v_{l_3}}\frac{a_{l_1 m_1}a_{l_2 m_2}a_{l_3 m_3} - 3\,C_{l_1 m_1 ,l_2 m_2}a_{l_3 m_3}}{\sqrt{C_{l_1}C_{l_2}C_{l_3}}}R_{n l_1 l_2 l_3}\) \right. \\
&\qquad \qquad\times \left.\(\frac{\curl{G}^{l'_1 l'_2 l'_3}_{m'_1 m'_2 m'_3}}{v_{l'_1}v_{l'_2}v_{l'_3}}\frac{a_{l'_1 m'_1}a_{l'_2 m'_2}a_{l'_3 m'_3} - 3\,C_{l'_1 m'_1 ,l'_2 m'_2}a_{l'_3 m'_3}}{\sqrt{C_{l'_1}C_{l'_2}C_{l'_3}}}R_{n l'_1 l'_2 l'_3}\)\>\\
&= \sum_{l_i m_i l'_i m'_i}\frac{\curl{G}^{l_1 l_2 l_3}_{m_1 m_2 m_3}\curl{G}^{l'_1 l'_2 l'_3}_{m'_1 m'_2 m'_3}}{v_{l_1}v_{l_2}v_{l_3}v_{l'_1}v_{l'_2}v_{l'_3}} R_{n l_1 l_2 l_3}\left[ 6 \, \langle a_{l_1m_1}a_{l'_1m'_1} \rangle \langle a_{l_2m_2}a_{l'_2m'_2}\rangle \langle a_{l_3m_3}a_{l'_3m'_3}\rangle \right.\\ 
\nn &+ \;9 \, \langle a_{l_1m_1}a_{l_2m_2}\rangle\langle a_{l'_1m'_1}a_{l'_2m'_2}\rangle\langle a_{l_3m_3}a_{l'_3m'_3}\rangle- 9\, C_{l_1m_1,l_2m_2} \langle a_{l'_1m'_1}a_{l'_2m'_2}\rangle \langle a_{l_3m_3}a_{l'_3m'_3}\rangle \\
\nn &-\left.\;9 \,\langle a_{l_1m_1}a_{l_2m_2}\rangle C_{l'_1m'_1,l'_2m'_2} \langle a_{l_3m_3}a_{l'_3m'_3}\rangle  + 9 \,C_{l_1m_1,l_2m_2} C_{l'_1m'_1,l'_2m'_2} \langle a_{l_3m_3}a_{l'_3m'_3}\rangle + ...\right] R_{n l'_1 l'_2 l'_3}\\
\nn &= 6 \sum_{l_i m_i l'_i m'_i}\frac{\curl{G}^{l_1 l_2 l_3}_{m_1 m_2 m_3}\curl{G}^{l'_1 l'_2 l'_3}_{m'_1 m'_2 m'_3}}{v_{l_1}v_{l_2}v_{l_3}v_{l'_1}v_{l'_2}v_{l'_3}} R_{n l_1 l_2 l_3} \frac{C_{l_1 m_1 ,l'_1 m'_1}C_{l_2 m_2 ,l'_2 m'_2}C_{l_3 m_3 ,l'_3 m'_3}}{\sqrt{C_{l_1}C_{l_2}C_{l_3}C_{l'_1}C_{l'_2}C_{l'_3}}} R_{n' l'_1 l'_2 l'_3} + \hbox{higher order}\\
&= 6 {\cal R\,C\,R}^T = 6 \zeta_{nn'}\,.
\end{align}

Our inverse covariance weighted estimator  \eqref{eq:newmodalestimator}  becomes 
\begin{align}\label{eq:modalest}
{\cal E} = \frac{\a_n \zeta^{-1}_{nn'} \b_{n'}}{\a_n \zeta^{-1}_{nn'} \a_{n'}}\,.
\end{align}
This result applies equally to the general anisotropic case where we have supplemented with basis functions $\curl{R}^{\,l_1 \,l_2\, l_3}_{m_1 m_2 m_3}$ which cannot be simply represented in terms of the Gaunt integral; the derivation above is unaffected.   Using \eqref{eq:modalest}, we note that the variance $\sigma^2$ of the optimal modal estimator \eqref{eq:modalest} becomes simply 
\begin{align}\label{eq:modalvariance}
\sigma^2 =  \frac{\a_n \zeta^{-1}_{nn'} \langle \b_{n'} \b_{p'}\rangle \zeta^{-1}_{pp'}\a_p  }{(\a_n \zeta^{-1}_{nn'} \a_{n'})^2} = \frac{6}{\a_n \zeta^{-1}_{nn'} \a_{n'}}\,.
\end{align}

\subsection{Bivector representation of covariance matrix}

Further efficiencies are made possible by identifying the underlying bispectrum `noise' and `mask' shapes that 
contribute to the covariance matrix. We can identify such a shape because the noise is correlated in 
harmonic space.    Empirically in our previous isotropic WMAP analysis we have found that the noise
and mask contribute in tandem to produce a single shape vector $\beta = \{\tilde\beta_{n}\}$ which we could identify  \cite{Fergusson:2010dm}. This noise/mask vector is significantly correlated with local non-Gaussianity which explains the difficulty in obtaining optimal constraints on this model.    The dominant effect of $\beta$ leads to the remarkable result that we can accurately represent the isotropic covariance matrix simply as the identity matrix perturbed by this bivector, that is,  
\begin{align}\label{eq:ic_bivector}
\z_{n n'} ~\px~ \delta_{nn'} + \tilde\beta_n\tilde \beta_{n'}\,.
\end{align}
For this, there is a  simple inverse covariance
\begin{align}\label{eq:ic_bivectorinv}
\z_{n n'}^{-1} ~\px~  \delta_{nn'} -  \frac{\tilde\beta_n\tilde \beta_{n'}}{1+ |\tilde\beta|^2 }\,, \qquad \hbox{where}\quad|\tilde\beta|^2  \equiv \sum_{i=1}^{\nmax} {  \tilde\beta_i}^2\,.
\end{align}
For the isotropic estimator \eqref{eq:modalest} in the presence of this noise/mask bivector the variance from  \eqref{eq:modalvariance} becomes 
\begin{align}\label{eq:bivecvariance}
\sigma^2\; &= \;\frac{ 1}{|\a|^2}\left ( 1  + \frac{(\a\cdot\tilde\b)^2}{|\a|^2(1+|\tilde\b|^2) - (\a\cdot\tilde\b)^2}\right) \nn\\
&=  \;\frac{ 1}{|\a|^2}\left ( 1  \;+\;\frac{|\tilde\b|^2\,\cos ^2\theta }{1 + |\tilde\b|^2\,(1- \cos^2\theta)}\right) \,,
\end{align}
where we have described the degree of correlation between the two vector directions through the cosine
\begin{align}
\a\cdot\tilde \b = \sum_{i=1}^{\nmax} \a_n\tilde\b_n \equiv |\a||\tilde\b| \, \cos \theta\,.
\end{align}

From \eqref{eq:bivecvariance}, we see that the variance is made up of two components.  First, there is the ideal or `optimal' variance $1/|\a|^{2}$ that can be achieved in the presence of the underlying Gaussian CMB signal and homogenous instrument noise, given the degree of sky coverage and beam resolution (incorporated into the signal-to-noise definition of $\a$).   Secondly, there is the contribution from the inhomogeneous noise and the anisotropic mask which arises if the noise/mask vector $\tilde \b$ is closely aligned with the model $\a$ under investigation.   Significantly, this second contribution vanishes if the model being studied has little correlation with $\tilde \b$, in which case near-optimal variance can be approached even in this isotropic case. In principle, in the full anisotropic case, the general inverse covariance weighting will approach optimality given sufficient modes for any model. 

Another key point in a practical implementation  is that the single noise/mask vector $\tilde\beta$ can be estimated accurately from relatively few simulations.   Using $\tilde \beta$ with the covariance matrix in the form (\ref{eq:ic_bivector}) allows accurate and efficient inversion because we only need to determine $\tilde \b$. This can be achieved with $n_{\rm sim}\lesssim 1000$ simulations, while direct averaging of the covariance matrix takes $n_{\rm sim}\px 10000$ to converge to percent levels (making it reliably invertible).  We remark that early studies of these contributions in a WMAP-realistic context showed that the local shape is strongly correlated with the spurious contributions from the mask and inhomogenous noise, with the noise/mask vector $\tilde \b$ showing an 70\% correlation with the local $\a^{\rm loc}$ direction.   Separately, there was about an 70\% local-mask correlation and a 47\% local-noise correlation.  In a forthcoming publication, we shall discuss the more general case where several `noise' vectors can be obtained using a principal component analysis (PCA), separating the noise, mask and other contaminant vectors from model directions of theoretical interest. 

Finally, we note that for the simple bivector form of the inverse covariance matrix \eqref{eq:ic_bivectorinv}, the Cholesky decomposition of $\z$ can also be expressed analytically.  Here, we have $\z = \tilde \lambda{\tilde\lambda}^T$ with $\tilde\l$ the lower triangular matrix:
\begin{align}
\tilde\lambda = \left( \begin{array}{cccc}
  {\textstyle{\sqrt{1+ {  \tilde\beta_1}^2}}} & 0 &0& 0 \\
 \frac{  \tilde\beta_2 \tilde\beta_{1'}}{\sqrt{ 1+ {  \tilde\beta_1}^2}}& \frac{\sqrt{1+ {  \tilde\beta_1}^2+{  \tilde\beta_2}^2}}{ \sqrt{1+ {  \tilde\beta_1}^2}}& 0& 0\\
...& ... & ...&...\\
\frac{  \tilde\beta_n \tilde\beta_{1}}{\sqrt{ 1+ {  \tilde\beta_1}^2}}& ... & \frac{  \tilde\beta_n \tilde\beta_{n'}}{\sqrt{ { 1+ \sum_{1}^{n'-1} {  \tilde\beta_n}^2}}\sqrt{ { 1+ \sum_{1}^{n'} {  \tilde\beta_i}^2}}} &  \frac{\sqrt{1+ \sum_{1}^n{ \tilde\beta_i}^2}}{ \sqrt{1+ \sum_{1}^{n-1} {  \tilde\beta_i}^2}}
\end{array}
\right) \,. 
\end{align}
Rotation with this matrix $\tilde \l$ using \eqref{eq:Cholrot}, offers an orthonormal frame in which the bispectrum estimator \eqref{eq:modalest} becomes simply ${\cal E} = \a' \cdot \b' /\,|\a'|^2$, that is, the effects of the noise and mask are incorporated into the definitions of $\a'$ and $\b'$ from the outset. 

\subsection{Isotropic inverse covariance implementation}

The practical implementation of the inverse covariance weighting in  \eqref{eq:newmodalestimator} has been incorporated in the isotropic case for the modal bispectrum estimator.   Bispectrum constraints for a wide variety of models using WMAP data have been presented elsewhere \cite{Fergusson:2010dm}.  Here, our purpose is to focus on the improvement in optimality for the variance $\Delta \fnl$ and we only present results for local,  equilateral  and orthogonal non-Gaussian models.   We compare the variance achieved for the modal isotropic estimator (which agrees well with previous isotropic KSW estimators)  to show that it can be improved with modal inverse covariance weighting in this isotropic case.  Of course, we do not achieve full optimality, which can be seen by comparing to the Fisher matrix forecast, because relevant anisotropic effects deriving chiefly from the mask cannot be described using only an isotropic basis.   

For the purposes of direct comparison, we present the variance $\Delta \Fnl$ using the integrated bispectrum $\Fnl$ defined in ref.~\cite{Fergusson:2010dm}, that is, the total bispectrum signal-to-noise normalised relative to the local model (for which $\fnl^{\rm loc} = \Fnl$).    For an arbitrary predicted theoretical bispectrum $b^{\rm th}_{l_1 l_2 l_3}$, we use the estimator \eqref{eq:bispectrumest} to define
\eq \label{eq:newfnl}
\Fnl^{\rm th} = {\cal E}, \quad\mbox{with}\quad N^2 \equiv  N_{\rm th}N_{~\rm loc}^{\fnl=1} \,,
\qe
where the universal normalization $N$ has been defined using 
\eq\label{eq:theorynorm}
{N_{\rm th}}^2 \equiv \sum_{l_i} \frac{\hlll^2{b^{\rm th}_{l_1 l_2 l_3}}^2}{C_{l_1}C_{l_2}C_{l_3}}\,.
\qe 
with $N_{~\rm loc}^{\fnl=1}$ being exactly the same quantity except defined for the $\fnl=1$ local model.  
The quantity $\Fnl$ provides a much more uniform variance between different models, though we also present results with the usual equilateral normalisation $\fnl^{\rm equil}$.

\begin{table}[h]
\begin{tabular}{| l | c | c | c | }
\hline
Estimation type & ~ Local $\Delta\Fnl ~ $ & Equilateral $\Delta\Fnl $ ($\Delta\fnl^{\rm equil}$)& Orthogonal $\Delta\Fnl $ ($\Delta\fnl^{\rm ortho}$)\\
\hline
{\bf Standard isotropic estimator } & $29.5\; $ & $ 24.1 \quad(129.0)$ & $ 26.0 \quad(107.9)$\\
{\bf Weighted isotropic estimator} & $27.6 \; $ & $ 23.8 \quad(127.4)$ & $ 25.0 \quad(103.8)$\\
{\bf Fisher matrix `optimal'} & $22.9$ & $22.9 \quad(122.5)$ & $ 22.9 \quad(94.9)$\\
\hline
\end{tabular}
\caption{Results for the variance from an implementation of the isotropic modal estimator in a WMAP realistic context.   These results were obtained from 1000 Gaussian maps at $\lmax = 500$ using the KQ75 mask and the WMAP inhomogeneous noise model.   The mask and noise degrades the variance relative to the Fisher matrix forecast, but this is significantly improved using modal inverse covariance weighting, even for this isotropic estimator.}
\label{tab:variance}
\end{table}

Results for the variance $\Delta \Fnl$  from local and equilateral models are listed in Table~\ref{tab:variance}.   This work conforms exactly to the WMAP-realistic methods described in ref.~\cite{Fergusson:2010dm}, using a KQ75 mask, WMAP beams and inhomogeneous noise model (see below) and with a resolution limit of $\lmax =500$.  The results were obtained using 144000 Gaussian maps from a modal bispectrum pipeline previously confirmed to be unbiased and which reproduces the correct $\fnl$ for simulated non-Gaussian maps \cite{FergussonLiguoriShellard2009}. Previous results for ideal full-sky Gaussian simulations (without inhomogenous noise or mask) demonstrate that the estimator is correctly normalised, confirming that optimality variance in the absence of experimental effects.   For the local model in a WMAP-realistic context, the variance from the modal estimator is suboptimal by 30\% dropping from the Fisher matrix forecast $\Delta  \Fnl = 22.9$ to a realised value $\Delta  \Fnl = 29.5$.  This is consistent with the value obtained from KSW estimators previously at the same $\lmax =500$ (e.g.\  $\Delta  \Fnl \px 26$ in ref.~\cite{WMAP5}).   This strong degradation for the local model occurs because the noise/mask  vector  \eqref{eq:ic_bivector})  correlates strongly with the local shape, increasing the variance as discussed previously (see \eqref{eq:bivecvariance}).   The addition of inverse covariance weighting for the local model, improves the variance to $\Delta  \Fnl = 27.6$ which takes us 30\% closer to the projected optimal variance, demonstrating the value of incorporating this weighting even in the purely isotropic case.   In contrast, nearly optimal results can always be obtained for the equilateral model and the orthogonal model because these are much less correlated with the noise/mask vector $\tilde\b$.  For the equilateral model, in particular, an isotropic estimator will be capable of achieving near-optimal results.

\section{Other polyspectra and 3D applications}
This methodology can be extended to other correlators. We begin with the trispectrum estimator presented in \cite{Regan:2010cn}, for which WMAP constraints were obtained for several models in \cite{Fergusson:2010gn}. We shall restrict ourselves to the diagonal free trispectrum for simplicity, although obtaining expressions for the general case is straight forward. We build the optimal modal estimator our of the quantities $\cal A,\, B,\, C$ in (\ref{eq:legoestimator}) defined by
\begin{align}
{\cal A} &=  \frac{G^{l_1 l_2 l_3 l_4}_{m_1 m_2 m_3 m_4} t_{l_1 l_2 l_3 l_4}}{\sqrt{C_{l_1}C_{l_2}C_{l_3}C_{l_4}}}\,,\\
{\cal B} &= \frac{a_{l_1m_1}a_{l_2m_2}a_{l_3m_3}a_{l_4m_4} - 6 \,C_{l_1m_1,l_2m_2} a_{l_3m_3}a_{l_4m_4} + 3\, C_{l_1m_1,l_2m_2}\,C_{l_3m_3,l_4m_4}}{\sqrt{C_{l_1}C_{l_2}C_{l_3}C_{l_4}}}\,,\\
{\cal C} &= \frac{C_{l_1 m_1 ,l'_1 m'_1}C_{l_2 m_2 ,l'_2 m'_2}C_{l_3 m_3 ,l'_3 m'_3}C_{l_4 m_4 ,l'_4 m'_4}}{\sqrt{C_{l_1}C_{l_2}C_{l_3}C_{l_4}C_{l'_1}C_{l'_2}C_{l'_3}C_{l'_4}}}\,,\\
{\cal R} &= \frac{G^{l_1 l_2 l_3 l_4}_{m_1 m_2 m_3 m_4}}{v_{l_1}v_{l_2}v_{l_3}v_{l_4}} R_{n l_1 l_2 l_3 l_4}\,,
\end{align}
where $v_l$ is a suitable weight function and the isotropising factor $G^{l_1 l_2 l_3 l_4}_{m_1 m_2 m_3 m_4}$ is
\begin{align}
G^{l_1 l_2 l_3 l_4}_{m_1 m_2 m_3 m_4} = \int Y_{l_1 m_1} Y_{l_2 m_2} Y_{l_3 m_3}Y_{l_4 m_4} = \sum_{LM} (-1)^M G^{l_1 l_2 L}_{m_1 m_2 M} G^{l_3 l_4 L}_{m_3 m_4 -M}\,.
\end{align}
Using efficient modal methods calculation of the trispectrum estimator with inverse covariance weighting is no more challenging than in the bispectrum.

Inverse covariance weighting for power spectrum estimation for high resolution experiments like Planck. In this case we would be constraining small deviations from a canonical model with a given $C_l$. Considering an isotropic subspace, the building blocks for the model estimator are
\begin{align}
{\cal A} &= \frac{\d_{l_1 l_2} \d_{m_1 m_2} \(\tilde{C}_{l_1}-C_{l_1}\)}{\sqrt{C_{l_1}C_{l_2}}}\,,\\
{\cal B} &= \frac{a_{l_1 m_1}a^*_{l_2 m_2} - C_{l_1 m_1 ,l_2 m_2}}{\sqrt{C_{l_1}C_{l_2}}}\,,\\
{\cal C} &= \frac{C_{l_1 m_1 ,l'_1 m'_1}C_{l_2 m_2 ,l'_2 m'_2}}{\sqrt{C_{l_1}C_{l_2}C_{l'_1}C_{l'_2}}}\,,
\end{align}
where $\tilde{C}_l$ is understood to be the canonical model including the predicted deviation. Also we now consider $C_{l_1 m_1 ,l_2 m_2} = \<a_{l_1 m_1}a^*_{l_2 m_2}\>$. The corresponding isotropic basis functions are
\begin{align}
{\cal R} = \frac{\d_{l_1 l_2} \d_{m_1 m_2}}{v_{l_1}v_{l_2}} R_{n l_1}\,,
\end{align}
which must be supplemented to incorporate anisotropic experimental effects such as cut-sky masking.   
We can exploit the tractability of the modal inverse covariance weighting in several ways. First we could consider constraining specific models with specific signatures in the power spectrum with a particular $\a_n$, such as oscillatory models. Secondly we could use the method for full power spectrum estimation using the recovered $\b$ to iteratively improve the best fit theoretical model. This would entail linking to Monte Carlo simulations which explore cosmological parameter space. Finally, we need not restrict ourselves to isotropic models and instead using additional anisotropic basis functions, we can to constrain anisotropy in the CMB.

For large-scale structure and other 3D polyspectra estimation, this approach has an entirely analogous implementation discussed in ref. \cite{Fergusson:2010ia}. We simply replace our multipoles with the density perturbation so now we have
\begin{align}
\langle \mathfrak{a}_{\boldsymbol \wp}\rangle ~\equiv~ \langle  \delta_{{\bf k}_1} \delta_{{\bf k}_2}... \delta_{{\bf k}_n}  \rangle 
\end{align}
where now the vector $\boldsymbol \wp$ represents ${\boldsymbol \wp} =(\bk_1,\bk_2,...,\bk_n)$ which cover the full $3n$-dimensional domain of possible polyspectra. Here momentum conservation and isotropy imply we can make the reduction to the subspace $ \wp = (k_1,k_2,..k_n)$ and the two are related by
\begin{align}
\mathfrak	{a}_{\boldsymbol \wp} = (2 \pi)^3 \delta ({\bf k_1} + {\bf k_2} + ... +{\bf k_n})\, \mathfrak{a}_{\wp}\,,
\end{align}
The rest of the modal method then follows analogously with two main caveats. The first being that the growth of structure is non-linear so even with gaussian initial conditions we produce measurable polyspectra.  The form of these can be calculated from $N$-body simulations and we can seek deviations from these due to primordial and other effects. Secondly there is a non-linear relationship between the late time and primordial polyspectra and so a Monte Carlo approach is required to find the parameter values which provide the best fit to the data. Error bars can then be obtained from simulations with the chosen primordial spectrum. However, the inverse covariance matrix can still be calculated from Gaussian simulations, so it does not need to be recalculated for differing levels of primordial non-Gaussianity.

\section{Conclusion}

We have shown that full inverse covariance weighting can be naturally incorporated in the 
modal polyspectra estimator to achieve optimality.   The resulting modal covariance matrix 
can be efficiently calculated and inverted.  This greatly reduces the computational cost and memory 
requirements of polyspectra estimation, making optimal analysis of high resolution experiments 
like Planck possible.    We emphasise that the primary efficiencies discussed
here arise because we have effectively projected out  all but the modal degrees of freedom needed
to characterise the model (or models) under investigation, while retaining an adequate description 
of the noise and systematics with which it is correlated. 

We have only briefly discussed the many applications of this work.  These include 
improved WMAP7 constraints on a wide variety of theoretical models, together with a model-independent 
constraint on the overall magnitude of the bispectrum in ref.~\cite{Fergusson:2010dm}.   A principal component 
analysis can be used to identify the noise, mask and other shapes for preconditioning the inverse 
covariance weighting.  A forthcoming paper will use eigenmodes to characterise and marginalise over foregrounds and 
other contaminants  in preparation for the CMB analysis of Planck data.    The analysis of higher order correlators in three dimensions represents
a particular challenge as many billion galaxy surveys become available.    The 3D modal estimator \cite{Fergusson:2010ia} allows for 
efficient polyspectra estimation and  for the setting of arbitrary non-Gaussian initial conditions.   Combined with inverse covariance weighting, 
this methodology may help unravel  the many 
systematic and evolutionary effects present in large-scale structure observations.

\section{Acknowledgements}

We are very grateful for many useful conversations with Michele Liguori and Donough Regan with whom we 
have developed the modal bispectrum methodology described here.   The inverse covariance weighting aspect of the work has been taken forward with Helge Gruentjen 
and we are very grateful for numerous enlightening discussions.    We also thank  Anthony Challinor for useful comments on a earlier draft of this paper   We are also grateful for 
helpful conversations and interactions with Eugene Lim,  Hiro Funakoshi, Marcel Schmittfull and 
Xingang Chen. 

Simulations were performed on the COSMOS@DiRAC supercomputer (an SGI Altix UV1), and the national DiRAC facility is funded by 
STFC and DBIS.  JRF and EPS were supported by STFC grant ST/F002998/1 and the
Centre for Theoretical Cosmology. 

\appendix

\section{Anisotropic noise and bispectrum optimality}

Anisotropic instrument noise provides an explicit example which illustrates some of the 
points raised above, notably the nature of its contribution to the variance the importance of the linear term in (\ref{eq:stdestimator}) for 
optimal estimation.  WMAP 
non-Gaussian analysis to date has used a simple instrument noise model which assumes no 
correlations between pixels.   If $n_{lm}$  is the harmonic transform of the pixel-variance
map, then the covariance of the noise contribution to the CMB temperature is given by 
\begin{align}\label{eq:anisotropicn}
\langle a^{\rm N}_{l_1m_1}a^{\rm N}_{l_2m_2}\rangle = \frac{4\pi}{n_{\rm pix}}\sum _{lm}(-1)^m \curl{G}^{\,l_1~l_2\;~ l}_{m_1 m_2 -m}\, n_{lm}\,.
\end{align}
(See, for example, the discussion of covariance under rotations in \cite{Hanson:2010gu}.)   
The covariance of the cut-sky mask also can be represented in this form by taking the masked pixels to have a very large noise so that the covariance weighting neglects that section of the map.
We shall now consider the contributions to the variance of the estimator both with and without the linear term. 

We assume CMB signal plus instrument noise can be represented as $a_{lm} + a^{\rm N}_{lm}$ with respective 
covariances $ C^{\rm S} _{l_1m_1,l_2m_2} = \langle a_{l_1m_1}a_{l_2m_2}\rangle = (-1)^{m_1} C_{l_1} \,\delta_{l_1 l_2}
\delta_{m_1-m_2}$ and $ C^{\rm N} _{l_1m_1,l_2m_2} =\langle a^{\rm N}_{l_1m_1}a^{\rm N}_{l_2m_2}\rangle$.   Again restricting attention to the isotropic subspace by expanding using isotropic modes \eqref{eq:isobasisbi}, the modal covariance matrix with the inhomogeneous noise takes the form 
\begin{align} \label{eq:pairvar}
\z_{nn'} ~~ &=~~{ \textstyle\frac{1}{6}} \sum_{l_i m_i l'_i m'_i}R_{n\,l_1 l_2l_3}\,\curl{G}^{\;l_1 \;l_2\; l_3}_{m_1 m_2 m_3}C_{l_1 m_1 ,l'_1 m'_1}C_{l_2 m_2 ,l'_2 m'_2}C_{l_3 m_3 ,l'_3 m'_3}\, \curl{G}^{\;l'_1\; l'_2 \;l'_3}_{m'_1 m'_2 m'_3}\, R_{n'l'_1 l'_2l'_3}\cr
\nn &=~~ { \textstyle\frac{1}{6}} \sum_{l_i m_i l'_i m'_i}R_{n\,l_1 l_2l_3}\curl{G}^{\;l_1 \;l_2\; l_3}_{m_1 m_2 m_3}\left (C^{\rm S}C^{\rm S}C^{\rm S} + 3 C^{\rm S}C^{\rm S}C^{\rm N} + 3 C^{\rm S}C^{\rm N}C^{\rm N}+ C^{\rm N}C^{\rm N}C^{\rm N}\right)  \curl{G}^{\;l'_1\; l'_2 \;l'_3}_{m'_1 m'_2 m'_3}\, R_{n'l'_1 l'_2l'_3}\cr
 &=~~ { \textstyle\frac{1}{6}}\, \sum_{l_i  l'_i }R_{n\,l_1 l_2l_3}\, R_{n'l'_1 l'_2l'_3}\left[ h_{l_1l_2l_3}^2 \,\delta _{l_1,l'_1} \delta _{l_2,l'_2} \delta _{l_3,l'_3}\left( C_{l_1}C_{l_2}C_{l_3}  + C_{l_1}C_{l_2}\frac{n_{00}}{\sqrt{4\pi}} \right)    \right.   \\
&\qquad\qquad\qquad\qquad~~~~+~\sum_l  h_{l_1l_2l_3}h_{l_1l'_2l'_3}h_{l_2l'_2l}h_{l_3l'_3l}  \left \{\medspace 
\begin{array}{ccc}
  l_1 & l_2 & l_3 \\
 l & l'_3 & l'_2
\end{array}
\medspace \right\}\;C_{l_1}C^{\rm N}_{l}    \cr
 &\qquad\qquad\qquad\qquad~~~~\left.+~\sum_{l\,l' l''} h_{l_1 l_2 l_3} h_{l'_1 l'_2 l'_3} h_{l_1 l'_1 l} h_{l_2 l'_2 l'} h_{l_3 l'_3 l''}   \left \{\medspace 
\begin{array}{ccc}
  l_1 & l_2 & l_3 \\
  l'_1 & l'_2 & l'_3 \\
 l & l' & l''
\end{array}
\medspace \right\}\;b^{\rm N}_{l\, l' l''}     \right]    
\end{align}
where $C^{\rm N} _l$ represents the isotropic power from the pixel-variance map, 
\begin{align}
C^{\rm N} _l \equiv \sum_m \frac{(-1)^{m}}{2l+1} n_{lm}n_{l-m}\,,
\end{align}
$b^{\rm N}_{l_1 l_2l_3}$  is the isotropic part of its bispectrum,
\begin{align}
b^{\rm N}_{l_1 l_2l_3}
=    \sum_{m_i} \curl{G}^{\;l_1 \;l_2\; l_3}_{m_1 m_2 m_3} n_{l_1m_1}n_{l_2m_2}n_{l_2m_2}\,.
\end{align}   
Standard identities were used to convert summed products of four 
and five Wigner-$3j$ symbols into the Racah-$6j$ and Racah-$9j$ symbols respectively. 

The expression (\ref{eq:pairvar}), which yields that the isotropic covariance from anisotropic noise (\ref{eq:anisotropicn}), depends solely on the two- and three-point correlators of the pixel-variance map (i.e.\  the isotropic functions $C^{\rm N} _{l}$ and $b^{\rm N}_{l_1 l_2l_3}$).    The reduction to isotropy means that 
 the product form of the full covariance matrix ${\cal C}$ can be reduced from a
$\lmax ^6 \times \lmax ^6$ matrix to $\lmax ^3 \times \lmax ^3$ but this is still intractable.  Here, however, we have a modal covariance matrix of only $\nmax \times \nmax$ which can be easily inverted.   

Now consider the contribution from inhomogenous noise to the covariance matrix in the absence of a linear term in the estimator \eqref{eq:bbispect} but without the linear term.   In this case, the full covariance has the 
following additional cross-terms 
\begin{align}\label{eq:crossterms}
 \langle\b_n \b_{n'} \rangle_{\times} &\propto ~9 \sum_{l_i m_i l'_i m'_i}\,R_{n\,l_1 l_2l_3}\,\curl{G}^{\;l_1 \;l_2\; l_3}_{m_1 m_2 m_3}
 C_{l_1 m_1 ,l_2 m_2}C_{l_3 m_3 ,l'_3 m'_3}C_{l'_1 m'_1 ,l'_2 m'_2}\,\curl{G}^{\;l'_1\; l'_2 \;l'_3}_{m'_1 m'_2 m'_3}\, R_{n'l'_1 l'_2l'_3}\cr
\nn &= ~ \sum_{l_i m_i l'_i m'_i}R_{n\,l_1 l_2l_3}\curl{G}^{\;l_1 \;l_2\; l_3}_{m_1 m_2 m_3}\left [9 C^{\rm S}C^{\rm S}C^{\rm S} + 3 (6 C^{\rm S}C^{\rm S}C^{\rm N} + 3 C^{\rm S}C^{\rm N}C^{\rm S})\right.\cr
\nn &\qquad\qquad\qquad \left. + 3(3 C^{\rm N}C^{\rm S}C^{\rm N}+6 C^{\rm S}C^{\rm N}C^{\rm N})+ 9C^{\rm N}C^{\rm N}C^{\rm N}\right]\curl{G}^{\;l'_1\; l'_2 \;l'_3}_{m'_1 m'_2 m'_3} R_{n'l'_1 l'_2l'_3}\cr
\nn &= ~9 \sum_{l_i  l'_i }R_{n\,l_1 l_2l_3}\,R_{n'l'_1 l'_2l'_3}\,\left[ \frac{(2l_1+1)(2l'_1+1)}{4\pi} \delta _{l_1,l_2} \delta _{l'_1,l'_2} \delta _{0,l_3} \delta _{0,l'_3}
\(C_{l_1}  C_{l'_1} C_0 + 2 C_{l_1} C_0 \frac{n_{00}}{\sqrt{4 \pi}}+ C_{l_1}  C_{l'_1} \frac{n_{00}}{\sqrt{4 \pi}}\) \right.\cr
\nn &\qquad\qquad\qquad\qquad + \frac{\left(2 l_1+1\right) h_{l'_1l'_2l'_3}^2}{4 \pi}(-1)^{l_1}  \delta _{l_1,l_2} \delta _{0,l_3}C_{l_1}C^{\rm N}_{l'_3} + \frac{ 2h_{l_1l_2l_3}^2h_{l'_1l'_2l'_3}^2  }{(2 l_3+1)} \delta _{l_3,l'_3}C_{l_3} C^{\rm N}_{l'_3}\\
&\qquad\qquad\qquad\qquad + \left.\sum_l\frac{h_{l_1 l_2 l_3}^2 h_{l'_1 l'_2 l'_3}^2}{\left(2 l_3+1\right) \left(2 l'_3+1\right)}\, b_{l_3 l'_3 l}\right]
\end{align}
The first line of the resulting expression corresponds to a monopole so can be ignored.  The remaining two lines containing three cross terms do not in general vanish, but once again they contain only contributions from the isotropic parts of the 
pixel-variance map correlators  $C^{\rm N} _{l}$ and $b^{\rm N}_{l_1 l_2l_3}$.     Other effects which break rotational invariance in a similar manner, such as a cut-sky mask, can be treated similarly.

It is important to note that the existence of the non-zero cross-terms (\ref{eq:crossterms}) means that the minimum variance cannot be achieved without use
of the linear term as shown in (\ref{eq:linearterm}).

\bibliography{Optimal}

\end{document}